# Physical Coordinate as Eigen Values of Translation Operators in Non-commutative Spaces

Takeo Miura

Advanced Institute for Mathematical Science, Ltd.

1-24-4, Motojyuku, Higashimatuyama, Saitama 355-0063, Japan

e-mail: t_miura@msg.biglobe.ne.jp

## Abstract

We show two independent theoretical methods which lead to generate eigen values of a composite two-particles system associated with the non-locality of EPR effect which was experimentally proved by Aspect and others. One method is to use the quantum groups (Woronowicz and Podleś) and the other is to use the two-dimensional $\kappa$-Poincaré groups. The representation of Hilbert spaces of these groups has eigen functions and eigen values with real numbers on one dimensional space. The range of eigen values is real number $[0, \infty)$, so that the composite two particles system in quantum groups can macroscopicly exists and it has a nonlocal effect.

## 1. Introduction.

*Bell*'s 1964 demonstration[1] that realistic interpretations of quantum theory must be nonlocal required the use of inequalities now universally known as Bell inequalities. A proof of nonlocality without inequalities for a two particles system had been given by Hardy[2] and others.

 We demonstrate the nonlocality mostly by means of concepts and techniques of quantum groups developed on $C^*$-algebra, which were invented and developed by Woronowicz[3] and Podleś[4], and then we interpret the quantum spaces as a geometric representation of real spaces. There exists a Hilbert space in which a spin-singlet state of two-particles system are constructed on quantum spaces. This singlet state has nonlocality.

As another independent method, we use mostly a method of two-dimensional $\kappa$-Poincaré groups[5]. By this method, the positive real numbers as eigen value of coordinate in a Hilbert space are constructed.

 We emphasize the " raison d'être " for non-commutative spaces. In this article, however, we only mention one paper published by Doplicher, Fredenhagen and Roberts[6], in which they described that "… space-time as a manifold shoud break down at very short distances(of the Planck length). The uncertainty relations which appear due to gravitational effects at the Planck scale naturally lead to the conjecture about non-commutativity of the coordinates…".

## 2. A synthetic interpretation of Aharonov-Bohm effect and EPR correlations as the identical phenomenon within the same category each corresponding to "interfering alternative path".

The key concept of this interpretation is the complementarity between path and interference. In Aharonov-Bohm effect, there is an observable phenomenon in which a charged particle can display interference phenomena produced by magnetic field. In EPR correlations, for example, a pion at rest decays into two photons, and we can check and confirm two passes of these photons as particles by means of experiment. Both phenomena are unified to a generalized concept as the path-integral which was introduced originally by Feynman. Feynman defined "interfering alternative path and their sum" [7] in order to discriminate physical phenomena.

We demonstrated [8] the existence of non-commutative spaces in Aharanov-Bohm effect, and according to the complementality mentioned above, we must demonstrate that EPR correlations need also to be based on non-commutative spaces.

In the following sections, we show solutions to these problems..

## 3. A geometrical interpretation of the orbit of the two electrons which emitted from an atom and travel far away from each other.

Quantum sphere ($S_{qc}^2$ sphere) [4] being homogeneous space under $SU_q(2)$ action to quantum space $\mathbb{C}(SU_q(2))$ [3] is determined by parameters space. These two-electrons decayed are represented by algebra on quantum sphere, and the quantum entangled states of these composite two-electrons system lead to a spin-singlet state. Quantum Sphere $S_{qc}^2$ are homogeneous spaces of quantum SU(2) groups $SU_q(2)$. Quantum spheres are generalization of the standard two-dimensional sphere $S^2$ endowed with a classical right action of SU(2) [or SO(3)]. The unique smooth nondegenerate irreducible representations of $SU_q(2)$ are $d_l$, $\ell=0, \frac{1}{2}, 1, \cdots$. where $d_0$ is trivial and $d_1$ is equivalent to an irreducible component in the decomposition of the tensor product u ⊗ u. The elements $e_1, e_0, e_{-1}$ satisfy the relations, $e_i^* = e_{-i}$, (i=-1, 0, 1), $a_{lm} e_l e_m = \rho$ I ( I is unit matrix). $b_{lm,k} e_l e_m = e_k$, (k=-1, 0, 1), $\lambda$, $\rho \in \mathbb{R}$. The above $C^*$ algebra is generally represented by C($Xq_{,\lambda,\rho}$) [9]、 and one obtains the equation $S_{qc}^2 = X_{q, 1-q^2, (1+q^2)^2 q^{-2} c+1}$, (c ∈ $\mathbb{R}$) by isomorphic map.

In the next paragraph we describe the eigen values and eigen spaces of Hilbert spaces of $\mathbb{C}(S_{qc}^2)$.

### 3.1. The eigen states.

The C($S_{\mu c}^2$) has the irreducible representatios [4]、c∈ [0, ∞[, 0<|$\mu$| <1: Projector $\pi_{\pm}$ are defined on a Hilbert space with an orthonormal basis.

$$A_{\pm} f_k = \lambda_{\pm} \mu^{2k} f_k , \quad B_{\pm} f_k = c_{\pm}(c)^{1/2} f_{k-1} ,$$

$$c_\pm(k) = \lambda_\pm \mu^{2k} - (\lambda_\pm \mu^{2k})^2 + c, \quad \lambda_\pm = \tfrac{-1}{2} \pm (c + \tfrac{1}{2})^{1/2}, \ k=0, 1, 2, \cdots$$

$f_k$ are eigen functions for k.

### 3.2. The Selfadjoint Operator.

Now consider products of the generators. Because of the Clebusch-Gordan relation $u^1 \otimes u^1 \cong u^0 \oplus u^1 \oplus u^2$, there are injective intertwiners $G^\alpha \in \text{Mor}(u^\alpha, u^1 \otimes u^1)$ for $\alpha \in \{0, 1, 2\}$, [11].

Since the corepresentations $u^\alpha$ are irreducible, there are constants $\lambda_\alpha \in \mathbb{C}$ such that for $\alpha \in \{0, 1\}$ which gives relation for the generators:

$$\sum_{k,l} e_k e_l G_{kl,t} = \lambda e_t \quad (\lambda = \lambda_1) \qquad (1)$$
$$\sum_{k,l} e_k e G^0_{kl,0} = \rho I \quad (\rho = \lambda_0) \qquad (2)$$

These generators are P, $P_{-1}, P_0, P_1$ and are given in [4] as the left-hand sides of equations (2b, 2c, 2d, 2e). $\Gamma$ is C*-homomorphism which operates as $\Gamma: C(X) \to C(X) \otimes C(G)$、and $\Gamma P = P \otimes I$. P is 4-components vector of P, $P_i$.

So far $\rho$ and $\lambda$ are parameters, because the left-hand sides of the above equations (1), (2) are selfadjoint operators, and the selfadjoint operators have discrete eigen values in the finite dimensional Hilbert spaces. But because $\rho$ is continuous value ( continuous spectrum ), these Hilbert spaces belong to Von Neumann algebra Ⅱ-type or Ⅲ-type. These eigen values are spectra, and are different from the parameters of $S^2$ spaces. To make this point clear, we must first consider deformation of the Poincaré group, and must express the two-electrons with the subgroup(boost operators) of the two-dimensional $\kappa$ -Poincaré group, and must lead the continuous values of a distance between two-electrons.

### 3.3. Two particles system on a quantum sphere [12].

Similarly as in Podleś[4], one can define *-algebra $A_c$ of polynomials on Quantum sphere $S^2_{qc}$, $c \in [0, \infty]$, as universal *-algebra with unity I, generated by three elements $e_{-1}, e_0, e_1$ satisfying the relations in the above equations (1), (2), with the parameters $\lambda$, $\rho$ given before in the relations (7a) and (7b) of [4]. The completion of $A_c$ in $(\cdot | \cdot)$ is a Hilbert space denoted by $h = L^2(S^2_{qc})$.

Nonzero elements of h describe quantum-mechanical states of one(scalar) quantum particles on quantum sphere. Therefore a state of a system consisting of two particles on a quantum sphere is described by a nonzero elements of h⊗h (tensor product of Hilbert spaces). We also remark that Dirac Operator[10] can be defined on quantum sphere. We investigate the dynamics of two identical non-interacting particles on a quantum sphere. We assume that the dynamics of one particle on a quantum sphere is given by one-particle Hamiltonian H∈B(h) such that H*=H and $H(A_c) \subset A_c$. We define two-particle Hamiltonian $H^2 = H_1 + H_2$, where $H_1$ =H⊗id, $H_2$=K(H⊗id)$K^{-1}$.

We interpret K as the operator of interchanging of two identical particles on the quantum sphere[12].

Let one-particle Hamiltonian H be such that $He_i = \lambda_i e_i$, i=$\frac{-1}{2}, \frac{1}{2}$, $\lambda_{-\frac{1}{2}}, \lambda_{\frac{1}{2}}$ for some real numbers. Using (6)-(8),(10)-(11),(14)-(16) in [12]、one can derive that

$$v = e_{\frac{1}{2}} \otimes e_{-\frac{1}{2}} - q e_{-\frac{1}{2}} \otimes e_{\frac{1}{2}} \in h \otimes h.$$

The above two-particles is a composite system state and is a spin-singlet state, and this state is the entangled state.

### 3.4. The difference between $S_{1c}^2$ and $S^2$.

We explain the EPR effect by using $S_{qc}^2$ space which has some features of non-locality after all. On one side, isomorphism($S_{10}^2 \approx S^2$) is such that $e_0 = \cos\theta$, $e_\pm = \pm\frac{1}{2}i\sin\theta\, e^{\pm i\psi}$. On other side, all $S_{1c}^2$ (c > $-\frac{1}{4}$) are isomorphic so that we may put $e_0 = 2(c+\frac{1}{4})^{\frac{1}{2}}\cos\theta$, $e_\pm = \pm i(c+\frac{1}{4})^{\frac{1}{2}}\sin\theta\, e^{\pm i\psi}$, for the generators of $-\frac{1}{4} < c < \infty$. In orthogonal coordinates, they are $e_\pm = \pm i(x_1 + x_2)$, $e_0 = 2x_3$, We interpret $x_k$ as the Cartesian coordinates on $S_{qc}^2$ in some scale of length. In case of R=$\frac{1}{2}\rho^{\frac{1}{2}}$ ($\rho$ =4c+1), at q = 1, R is a radius of sphere $S^2$[4]. Thus the coordinate so called radius is the eigen value with positive real numbers. In the next paragraph we describe that a radius R of quantum sphere can be identified with one-dimensional distance R=R(t).

## 4. The explanation of R=R( t ) by the method of quantum space-time symmetry.

As the basis for the explanation, one must consider inhomogeneous groups under Poincaé transformation.

We first describe the order of our explanation in this section. At first step, we describe the $\kappa$ -Poincaré group、and then $\kappa$ -Poincaré algebra. Then we describe their duality.At second step, we describe the two-dimensional $\kappa$ -Poincaré group and two-dimensional $\kappa$ -Poincaré algebra, and then their duality, and they are inhomogeneous subgroups (boost operators).On these results, two-dimen- sional $\kappa$ -Poincaré quantum groups are described.

At last step, we consider the translation. Here two-dimensional dual $\kappa$ -Minkowski space is introduced. This two-dimensional space-time satisfies noncommutative relations. The representation of Hilbert space of one dimensional subgroup as translation has positive continuous values and these values are the coordinates.

### 4.1. $\kappa$ -Poincaré algebra.

The following theory depends on the use of [13] as basic frame and so called $\kappa$ -Poincarė algebra[5].

The algebra $P_\kappa$ has a natural bicrossproduct structure[13][14],
$$P_\kappa = T \triangleright\!\triangleleft U(SO(3,1))$$
T is defined as an algebra generated by $P_\mu$  $\mu=0,\cdots,3$ obeying the following relations:

$[P_\mu, P_\nu]=0$,   $\Delta P_0 = P_0 \otimes I + I \otimes P_0$,  $\Delta P_i = P_i \otimes e^{-\frac{P_0}{\kappa}} + I \otimes P_i$,

$S(P_\mu) = -P_\mu$,   $\varepsilon(P_\mu)=0$,

### 4.2. $\kappa$-Poincaré group[14].

The $\kappa$-Poincaré group $\tilde{P}_\kappa$ can be defined as a bicrossproduct:
$$\tilde{P}_\kappa = T^* \triangleright\!\triangleleft C(SO(3,1))$$
$C(SO(3,1))$ is the standard algebra of functions defined over Lorentz group while $T^*$ is defined by the following relations:

$[x^\mu, x^\nu]=\frac{1}{\kappa}(\delta_0^\mu x^\nu - \delta_0^\nu x^\mu)$  ,  $\Delta(x^\mu)=x^\mu \otimes I + I \otimes x^\mu$,

$S(x^\mu)=-x^\mu$,   $\varepsilon(x^\mu)=0$,

### 4.3. Duality[14].

We shall define the duality between the $\kappa$-Poincaré algebra and the $\kappa$-Poincaré group as follows:

$$C(SO(3,1)) \Leftrightarrow U(SO(3,1))$$
$$T^* \Leftrightarrow T ,$$

We define standard duality between the Lorentz group and algebra as follows:

$$<\Lambda_\nu^\mu、M_{\alpha\beta}> = i(\delta_\alpha^\mu g_{\nu\beta} - \delta_\beta^\mu g_{\nu\alpha}).$$

The duality $T^* \Leftrightarrow T$ is defined by $<x^\mu、P_\nu> = i\delta_\nu^\mu$.

### 4.4 The case of two dimensional $\kappa$-Poincaré algebra.

The deformational subgroup of the three dimensional rotational group of four dimensional deformation Poincaré group is represented by quantum sphere by Woronowicz and Podleś. The other subgroup ( boost ) is two-dimensional $\kappa$-deformational Poincaré group, and the decayed two-electrons run oppositely to each other on one straight line conserving momenta. Thus the translation is introduced.

### 4.5. Two dimensional $\kappa$-Poincaré algebra.

The algebra $F_{un\,\kappa}(P^2)$ can be constructed of two subalgebras, Fun(SO(1,1)) and $Fun_\kappa(T^2)$ making use their bicrossproduct.

The algebra Fun(SO(1,1)) is generated by the commuting elements $M_\nu^\mu$ ($\mu$, $\nu$=0,1) of 2×2 Lorentz matrix M with the natural Hopf structure.

$\Delta(M_\nu^\mu)=M_\rho^\mu \otimes M_\nu^\rho$,   $\varepsilon(M)=I$,

$S(M) = \eta^{-1} M^T \eta$,

where $\Delta$ is the commultiplication, $\varepsilon$ is the counity and S is the antipode: $\eta$ is the Minkowski metric tensor: $\eta$=diag{1,-1}.

The pseudo-orthogonality of SO(1,1) gives the relations $M^T \eta M = \eta$,

i.e. $(M_0^0)^2 - (M_1^0)^2 = 1$, $M_1^1 = M_0^0$, $M_1^0 = M_0^1$, so that Fun(SU(1,1)) has only one independent generator, e.g. $M_0^0$.

The algebra $Fun_\kappa(T^{(2)})$ is generated by two independent elements with the following relations,

$$[u^0, u^1] = i\kappa u^1, \quad \varepsilon(u^\mu) = 0.$$
$$\Delta(u^\mu) = u^\mu \otimes 1 + 1 \otimes u^\mu, \quad S(u^\mu) = -u^\mu.$$

It is easy to recognize in these relations the standard Hopf algebra structure of the undeformed Lie algebra $igl(1)$ of inhomogeneous transformations of real line.

### 4.6. Two dimensional $\kappa$-Poincaré group.

The complete Hopp algebra $P_\kappa^{(2)} = ISO_\kappa(1,1)$ is constructed as the bicrossproduct $P_\kappa^{(2)} = T_\kappa^{(2)} \triangleright\!\!\blacktriangleleft SO(1,1)$ with the help of structure map $\triangleleft$ : Fun(SO(1,1)) $\otimes Fun_\kappa(T^{(2)}) \to$ Fun(SO(1,1))、which read as follows,

$$M_0^0 \triangleleft u^0 = i\kappa((M_0^0)^2 - 1), \quad M_0^0 \triangleleft u^1 = i\kappa M_0^1(M_0^0 - 1),$$

And the coaction $\beta$ : $Fun_\kappa(T^{(2)}) \to$ Fun(SO(1,1))$\otimes Fun_\kappa(T^{(2)})$

$$\beta(u^\mu) = M_\nu^\mu \otimes u^\nu$$

The map $\triangleleft$, $\beta$ [13], in frame of the bicrossproduct construction give the following defining relations for the complete Hopp algebra $Fun_\kappa(P^{(2)})$

$$[u^0, M_0^0] = i\kappa((M_0^0)^2 - 1),$$
$$[u^1, M_0^0] = i\kappa M_0^1(M_0^0 - 1),$$
$$\Delta(u^\mu) = u^\mu \otimes 1 + M_\nu^\mu \otimes u^\nu, \quad S(u^\mu) = -\eta^{\mu\nu} M_\nu^\rho \eta_{\rho\lambda} u^\lambda$$

### 4.7. The Duality of Two-dimensional $\kappa$-Poincaré quantum group.

It is interesting to note that the quantum group $P_\kappa^{(2)} = ISO_\kappa(1,1)$ is constructed from the usual Lie group SO(1,1) and the usual Lie algebra $igl(1)$. Correspondingly the dual quantum universal enveloping algebra $U_\kappa(iso(1,1))$ is constructed from the Lie algebra $so(1,1)$ and Lie group IGL(1,$\mathcal{R}$).

### 4.8. dual two-dimensional $\kappa$-Minkowski space.

The non-commutative coordinates $x^0, x^1$ of the two-dimensional $\kappa$-Minkowski space $M_\kappa^{(2)}$ have the commutation relations similar to those of Parameters of translations $u^\mu$ of $P_\kappa^{(2)}$ $\quad [x^0, x^1] = i\kappa x^1$, i.e. from Lie algebra $igl(1)$.

### 5. The operators(translation and dilataion) of the two-dimensional $\kappa$-Minkowski space coordinate are noncommutative and have eigen values[5].

From the dicussed structure of $P_\kappa^{(2)}$, it is clear that the conjugate components of momenta parametrize the Lie group IGL(1,$\mathcal{R}$). Thus the $\kappa$-Minkowski space has the same general structure. The space-time coordinates are a Lie algebra and the conjugate components of momenta are parametrizing the corresponding Lie group. The representations of the IGL(1, $\mathscr{R}$) is constructed in the space $\mathscr{H}_{M_\kappa^{(2)}} = L^2(\mathscr{R}_V^{(2)})$ by quadratically integrable functions $f(g)=f(b,a), g \in$ IGL(1, $\mathscr{R}$) defined on upper half-plane $\{a>0, -\infty<b<\infty\}$, the domain of variation of the parameters of IGL(1, $\mathscr{R}$). Note that if IGL(1,$\mathscr{R}$) is considered as the group of translations on real line, the parameter 'a' corresponds to dilatations (which form multiplicative subgroup $\mathscr{R}_+$ of positive numbers), and 'b' corresponds to translations (which form additive subgroup $\mathscr{R}$).

Now we describe the eigen spectra and eigen functions of translation operator.

The (right)regular representation on $\mathscr{H}_{M_\kappa^{(2)}}$ is defined by the equality

$$R_{g_0} f(g) = f(gg_0)$$

where the group multiplication has the form $g(b_1, a_1)g(b_2, a_2) = g(b_1+a_1 b_2, a_1 a_2)$.

To decompose $\mathscr{H}_{M_\kappa^{(2)}}$ to the irreducible components, consider the subspaces $\mathscr{H}_{M_\kappa^{(2)}} \lambda \in \mathscr{R}$ of functions of the the form $f_\lambda(b,a) := \int dt\, e^{-it\lambda} f(b+t, a)$, with the property $f_\lambda(b+b_0, a) = e^{i\lambda b_0} f_\lambda(b,a)$.

It is easy to check that the subspaces $\mathscr{H}_{M_\kappa^{(2)}}$ are invariant with respect to transformations from IGL(1,$\mathbb{R}$). It is clear that for any $f_\lambda \in \mathscr{H}_{M_\kappa^{(2)}}$, one has

$f_\lambda(b,a) = e^{ib\lambda} f_\lambda(0,a) \equiv e^{ib\lambda} \phi(a)$, so that the irreducible representations $R_\lambda(g)$ of IGL(1,$\mathbb{R}$) is constructed in the space $\mathscr{H}_{M_\kappa^{(2)}}$ of functions $\phi$ on a half-line $0<\xi<\infty$ and has the form $\quad R_\lambda(g(b,a))\phi(\xi) = e^{\lambda b \xi} \phi(a\xi)$.

In fact these representations are induced by the one-dimensional representations of the translation subgroup $T^{(1)}$ of elements $g(b,1)$.

The infinitesimal operators corresponding to a representation $R_\lambda$ have the form

$\quad d_R = a\partial_a$ (generator of dilatations), $\quad t_R = a\partial_b$ (generator of translations).

As we discussed above, these infinitesimal operators, being the right-invariant vector fields on IGL(1, $\mathbb{R}$), play the role of non-commutative coordinates of configuration subspace. Thus we identify

$$x^0 = i\kappa d_R, \quad x^1 = i\kappa t_R .$$

One can immediately see that the quantum number $\lambda$, which distinguish different subspaces $\mathcal{H}_{M_\kappa^{(2)}}^\lambda$ is just the eigen value of $t_R$. Diagonalizing one of the right-invariant operators, e.g. $t_R$, so that

$$t_R \varphi_{\lambda,k}(\xi) = ik\lambda \varphi_{\lambda,k}.$$

Thus we have the basis of $\mathcal{H}_{M_\kappa^{(2)}}$, labeled by two numbers $\lambda, k \in \mathcal{R}$. From the point of view of quantum mechanics on IGL(1,$\mathbb{R}$) the functions $\varphi_{\lambda,k}(\xi)$ are eigen functions of the coordinate operators and the existence of the two quantum numbers(values of coordinates) corresponds to the two commuting components a and b of momenta. Thus we could prove our proposal R=R(t) by setting up R(t)≅λ(t).

## 6. Conclusion.

As for R described, if we suppose that R=R(t) depends on time, R(t) is the distance from the starting point where the two-electrons produced. Now $S^2$ is not quantum space, it has $\rho = 1$ (unit) which is a fixed value. On the contrary, noncommutative spaces can represent the nonlocality by the existence of R(t). Therefore EPR effect can be formed on noncommutative spaces. The quantum entanglement and the nonlocality can be represented geometrically on $S_{qc}^2$ but not on $S^2$. As a conclusion, EPR effect can be represented naturally and inherently on quantum groups. Also by means of two independent methods representing separately nonlocality, we have shown that the value of one-dimensional coordinate in Hilbert space is positive real numbers as eigen values.